# DEGASSING OF BIOLIQUIDS IN LOW ELECTROMAGNETIC FIELDS

Vladimir Shatalov*, Inna Noga, Alina Zinchenko

*Biophysical Department, Biological Faculty, Donetsk National University, Schorsa st., 46, Donetsk, 83050, Ukraine*

* Corresponding author. Tel: +380 50 2114531; E-mail: vladimir.shatalov@gmail.com

**Abstract**

A similarity of changes in physical-chemical properties of pure water induced by low electromagnetic fields (EMF) and by degassing treatment brought us to a conclusion that EMF produces some degassing of water. Degassing in turn gives rise to some biological effects by increasing the surface tension and activity of dissolved ions. In such a way the degassing can modify conformations of proteins and others biomolecules in bioliquids. That was confirmed in our observation of changes in the erythrocyte sedimentation rate and the prothrombinase activity in blood clotting processes.

## 1. Introduction

In the last years many investigations are devoted to the low electromagnetic field (EMF) that is producing non-ionizing and non-thermal effects in water or water systems. Water is the media of inhabitation and essential component of the majority of alive organisms. Nowadays this media is filled with anthropogenic electromagnetic fields. According to World Health Organization [1] and USA National Institute of Health [2] the fields can produce hardly predictable health after-effects. But the question weather low EMF affects human health is still open because there are no any proved theoretical model that could explain non-ionizing and non-thermal impact of EMF in poor conducting non-magnetic media like all alive are. Nevertheless EMF is applying in therapy without any theory and there exist widely discussing problems of water activated by EMF.

The most known sophisticated theoretical models were analyzed in [2]: effects of electric fields on cell surface structures and cell attachment, polarization forces, coulomb forces, effects of induced charge on cells, induced surface charge on cell matrix or substrate, cyclotron resonance and ion parametric resonance, Lednev's model, ion parametric resonance, biological electron transfer, effects on biogenic magnetite, magnetochemistry: effects of magnetic fields on free-radical reactions, non-linear dynamics and application of stochastic resonance. According to the report [2] all the models are controversial because, firstly, they contain unproved hypotheses about primary target of electromagnetic impact and subsequent biological effect. And secondly, these models are hardly verified in experiments.

Any theory solving the problem must answer at least the three main questions. What is: 1) the primary target, 2) the way of accumulation of the electromagnetic field impact, 3) the cause of biological effect? The aim the paper is to present a new model that gives the answers. The model is based only on facts observed in different experimental works and can be simply proved experimentally.

## 2. Exposure in low EMF and degassing of liquids produce the same effects

Let us analyze some facts first, without any hypothesis. It is well known that low electromagnetic fields change physicochemical properties of water like the redox potential, acidity pH, conductivity, turbidity [3, 4].

### 2.1. Conductivity and pH of pure water increase after EMF treatment or degassing

It was reported that the conductivity of pure water increased up to ten times and pH increased approximately on a unit after treatment for hours in a weak high-frequency EMF [4].The same results were obtained earlier by another author at another frequencies [5]. In [4] the observable changes had strongly pronounced frequency dependence with maxima nearby 170MHz. These data recalculated from [4] are presented in Figure 1. One can see that the linear transformation diminishes the reported frequency dependence. So the last might be caused by non-uniform conditions of the treatments.

Surprisingly, degassing of water acts in the same way [6, 7]. The observed increase of pH after degassing is well known and supposes the only explanation. It is caused by the carbon dioxide gas output and it means that other components of air also go out from water. Electrical resistance of degassed water is 0.8MΩ·cm (22°C) [7] comparing to 18MΩ·cm (25°C) for ultra pure water [8]. The authors believe that the reduction of electro resistance at degassing is caused by removal of the non polar gases $O_2$ and $N_2$ possessing structuring effects.

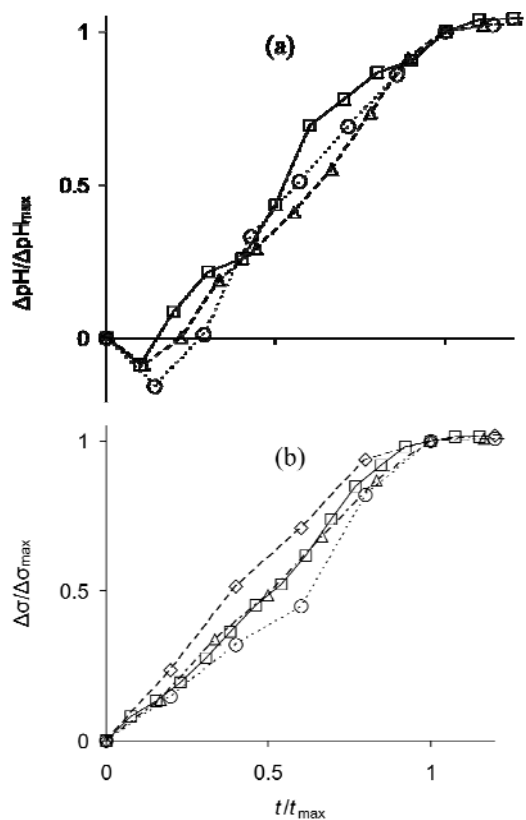

**Figure 1.** The plots of pH (a) and conductivity of water (b) changes in units of the maximum deviations via the irradiation time $t/t_{max}$ by low EMF of the different frequencies: 30, 110, 150 and 170MHz – circles, triangles, rhombi and squares correspondently (recalculated from [4])

## 2.2. Surface tension of pure water increases after EMF treatment or degassing

It was reported that an exposure in sufficiently high electric ~1MV/m [9] or magnetic ~10T [10] fields increase surface tension of water in 2%. Explanations given suppose these phenomena are caused by a stabilization of H-bonds or a dumping of surface waves. Degassing under EMF still never discussed but it gives the same result. Degassing also increases the surface tension of water up to 5% [6].

The investigation of micelle structuring processes under EMF in the surface-active substances solutions [11] showed changes in transparency and critical concentration of the micelle structuring. Last fact seems to be very important because the formation of micelles occurs under the same laws as formation of native structure of proteins. In such a way the changes of the properties of water under EMF may cause biological after-effects in bio-liquids.

## 2.3. The nature of after-effects

Many observed effects of treatment in fields have strongly pronounced after-effects. Some authors mention to these after-effects as a kind of "memory of water". The delays may be several hours or days long. Figure 2 shows a simple example that clear out the origin of such delays in water [18]. These after-effects of degassing look like produced by fields (Figure 1) and have a simple explanation. The degassing caused by coalescence of nanobubbles still continues after the field turns off until all the bubbles float up the surface. Then the absorption of air decreases pH to the equilibrium value.

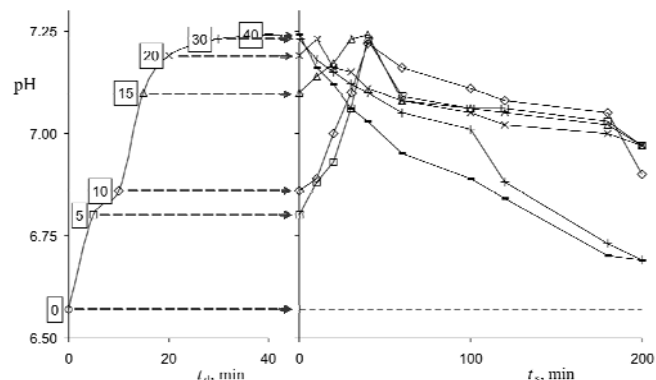

**Figure 2.** On the left – the plot of degassing, the dependence of water pH via centrifugation time $t_d$ displayed in the box near marker for each treated sample of water. On the right – the plot of air dissolution, pH relaxation after centrifugation to the initial value (the lower dashed curve) for each of the samples depending on the dissolution time $t_s$. The dashed arrows connect final points of centrifugation to starting points of air dissolution.

At this stage we have to state that water contacting air is an open non-homogenous system. The air solvated in water forms nano- and micro-bubbles. Velocity of the bubble floating up is proportional to its squared size. Changes in external conditions can stimulate a growth, merge and emersion of bubbles, that gives degassing of water. It takes a diffusion time somewhat about one hour or more. The paper [12] reports on a possibility of self organization of bubbles in pure water. The nanobubbles form clusters which can be seen in a trek of laser beam. So, electromagnetic fields produce degassing, and that is a true hypothesis. Now we have to find any biological effect of degassing.

## 3. Biological effects of degassing

### 3.1. Degassing increases hydrophobic forces

It is well known that the hydrophobic forces control the protein activity [13]. There exist experimental data [14-16] proving that EMF impacts the hydrophobic interaction in unknown way. These works proposed that biological effect of EMF is caused by conformations of biomolecules induced by changes of hydrophobic forces. The physics of the phenomena was not clear till now. If degassing changes the surface tension [6], another words – hydrophobicity, then the degassing can modify the protein activity like it happens in the micelle structuring [17]. So, if EMF produces degassing which changes hydrophobic interaction and modifies the protein activity then this is the direct biological effect of exposure in fields.

### 3.2. Degassing changes erythrocyte sedimentation rate

Another biological effect of degassing of bio-liquids concerns to the results of the erythrocyte sedimentation rate (ESR) tests after centrifugation [18]. The samples of blood of five different patients with initial values of ESR equal to 4, 7, 8, 9 and 17 mm/hour were treated in the 3000 cpm centrifuge for a number of time periods $t_d$. Figure 3 shows the averaged dependence of ESR relatively to initial value via the time of centrifugation of the blood samples – triangles. The curve with circles refers to ESR of preliminary extracted erythrocytes which then were posed in citrated plasma treated in the centrifuge. These results were averaged by samples obtained from three different patients with initial values of ESR equal to 5, 8 and 12 mm/hour. The both curves coincide within the standard deviation bars. So, the shape of observed dependences of ESR via $t_d$ is a result of the plasma degassing under centrifugation.

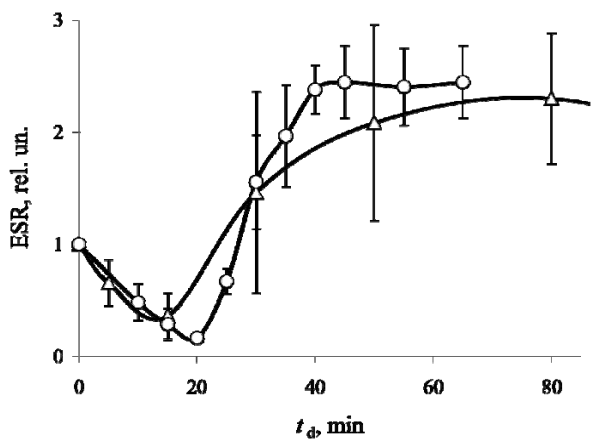

**Figure 3.** Erythrocyte sedimentation rate after $t_d$ minutes treatment in the 3000 c.p.m. centrifuge of the samples of blood (triangles) and plasma (circles) with the standard deviation bars and interpolation solid curves.

The degassing may modify erythrocyte sedimentation rate by the two ways. The first is via increasing viscosity of plasma which lows down the rate. The second is via increasing hydrophobic forces [6] which grow up the rate due to erythrocyte aggregations increase. The both, down and up trends, we can see at the plot. For the time exceeding the total degassing time (~50 minutes) the curves approach a stationary value which 2.5 times exceeds the initial one. That means the maximal aggregation of erythrocyte has 2.5 fold longer sizes.

### 3.3. Degassing changes the blood clotting time

The results of our research [19] of air dissolved in blood effect on dynamics of the blood clotting *in vitro* are presented at Figure 4. There exist two ways of blood clotting: internal way caused by a special protein, the blood prothrombinase, and external way caused by the tissue prothrombinase. The Figure 4 show the blood clotting time, or reverse activity of the proteins, as function of the time of centrifugation. We can see that centrifugation gives non monotonic changes of the activities. But the absorption of air while shaking the probe gives relaxation to the initial value of activity. Therefore, we can associate the centrifugation with degassing. And this is another biological effect of degassing.

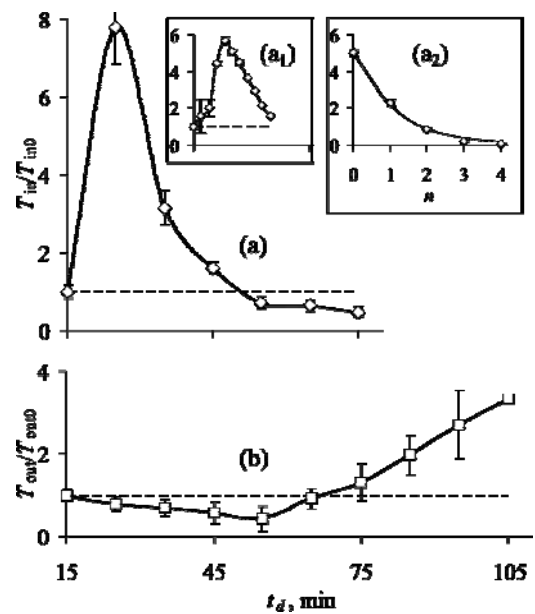

**Figure 4.** Internal (a) and external (b) blood clotting times in the units of correspondent initial values (dashed lines) as functions of the time of centrifugation td; the insert ($a_1$) shows in details the shape of maximum of $T_{in}(t_d)$ and the insert ($a_2$) presents the results of relaxation of the maximum value after absorption of air at the number n of 5-minutes shaking processing. $T_{in}$ were averaged over 4 and $T_{out}$ – over 3 samples of blood of different donors

### 3.4. Degassing increases Ca$^{++}$ activity in blood

The next biological effect of degassing is the most interesting and may be the most important. The Figure 5 shows the blood clotting time as a function of concentration of the calcium ions which initiate the blood clotting [19]. As it is seen from the plots the curves corresponded to a greater degassing time are shifted towards lower concentrations from 0.1 to 0.5mM. Such an increase of activity of the calcium ions in blood after degassing may be caused by removing the nanobubbles which are trapping the ions to get stability. So, EMF producing degassing of blood can impact on the dynamics of the blood clotting by increasing $Ca^{++}$ activity.

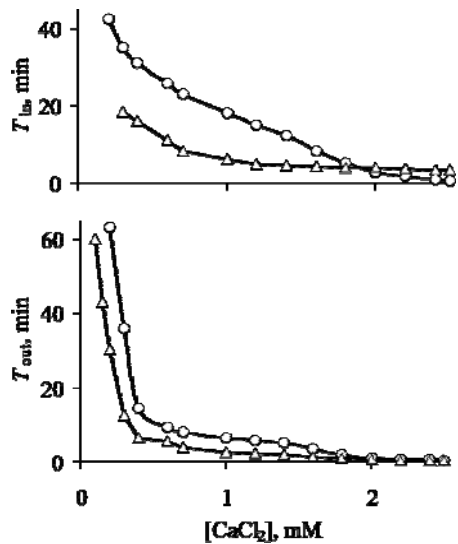

**Figure 5.** Internal clotting time $T_{in}$ after $t_d$ =15min – (o) and $t_d$ =60min – (Δ) centrifugation (top) and external clotting time $T_{out}$ after $t_d$ =15min – (o) and $t_d$ =60min – (Δ) centrifugation (bottom) in dependence on $CaCl_2$ concentration

So, if EMF produces degassing then it is the biological effect we are looking for. What is the physics of the phenomena?

## 4. The mechanism of non-thermal EMF bio-effect

The above results confirm that the colloid-water interface appears to be the most probable target of EMF. It was shown [20] that degassing of water can remove some observable EMF effects. The authors of [21] investigated EMF effects in the triply distilled and deionized water and supposed that the field may act on the gas dissolved in water. Effects of a pulsed low frequency EMF were investigated in the light scattering [22] and photoluminescence [23] of pure water to clear up the role of bubble-water interface. On the basis of rather careful experiments the authors came to a conclusion that EMF effects are lower in water with the reduced maintenance of dissolved air.

### 4.1. Bubbles trap ions to get stability

The dissolved gases usually are presented even in the distilled and deionized water partially in the form of micro or nanobubbles. In the equilibrium the processes of dissolution and degassing also as born and decay of bubbles compensate each other. The nucleation in water of micro and nanobubbles has, obviously, a fluctuation origin with a probability that is inversely proportional to a square root from volume of a bubble. Bubbles trap ions to get stability (as in Expansion or Bubble chambers in Nuclear Physics). Such charged bubbles should have a lower mobility comparing to mobility of hydrated ions. That is the reason why the conductivity increases after degassing (or exposure of water in electromagnetic field). The theory of the bubble charging was developed in [24]. The authors show that the charged bubbles (called bubstons) can form clusters of micron size. These bubstons were observed in the light scattering experiments [25]. So, the nanobubbles in water also as in bioliquids are really existing phenomena. EMF treatment just accelerates a coagulation or coalescence of nanobubbles that move them out from the liquid.

### 4.2. EMF induces bubble to bubble and bubble to border attraction

According to [26] EMF polarizes the bubbles and induces bubble to bubble and bubble to border attractions and this way accelerates its growth. The estimation of the coalescence time made in [26] took into account the dipole-dipole attraction of the synphase polarized bubbles. This time was sensitive to the concentration of bubbles and the field strength and did not depend on frequency. Unfortunately the obtained time for nano-scaled bubbles was much longer then the observed one. We believe that this discrepancy may be overwhelmed by taking into account a polarization of the counter-ion clouds which always surround the multi-charged bubbles according to [24]. EMF can displace the multi-charged cloud on a macro distance that increases the attraction between bubbles proportionally to a squared product of the charge to displacement. Besides, may be EMF produces synphase mechanical oscillations of the bubbles which give rise to the well known Bjerknes attraction in ultrasound waves.

## 5. Conclusions

So, despite of all efforts of different investigators [27] the mechanism of non-ionizing and non-thermal impact of EMF in water was not clear till now. Many authors believe these effects were caused by changes in water structure under EMF treatment. Degassing under EMF still never discussed despite the presence of air in every sample of water contacting the atmosphere. It is possible to find in the literature messages on effective degassing of water without essential rise in temperature by means of microwaves and even by illumination [28]. In [29] concentrations of the dissolved and free oxygen in water were defined by the nuclear magnetic resonance method. The measurements shown that $1m^3$ waters at 20°C contains near $2 \cdot 10^{15}$ bubbles with average radius ~ 20nm. After 30min irradiations the total volume of air in the bubbles decreased almost twice whereas the quantity of the dissolved oxygen decreased just on 12 % that corresponded to reduction of solubility by the accompanying heating of water on 4.5°C. The mechanism of this phenomenon was not discussed; however, it is obvious that the irradiation leads to an output of bubbles from water without appreciable changes of the solubility of air.

The above analysis confirms the model [26] supposed that nanobubbles of dissolved air in water or bioliquids are primary targets of EMF. EMF treatment of water or bioliquids accelerates the coalescence, growth in sizes and emersion of the bubbles out from the liquid. The degassing changes hydrophobic interaction and increases activity of dissolved ions. That gives rise to biological effects.

This theory explains the observed changes of water properties in EMF also as the observed changes in erythrocyte sedimentation rate and the prothrombinase activity in blood clotting processes.

Bubbles as the primary targets of the low electromagnetic field biological effect is the main output of the work. Besides, many mysterious experimental results may be easily explained now. The increase of pH after EMF treatment [4, 5, 21-23] may be explained by $CO_2$ output; the increase of conductivity [4, 5] is due to changes in type of carriers after degassing from charged nanobubbles to hydrated ions possessed higher mobility; changes in transparency of irradiated water [4,11] are due to the bubble average size growth under EMF; after-effects of EMF [3, 21-23, 30] are due to slow floating up of the integrated micro-bubbles. Now we can answer the abovementioned questions: 1) the primary target of EMF is a bubble as the liquid density discontinuity; 2) the way of accumulation of EMF impact is degassing of the treated liquid; 3) the biological effect is caused by changes of physical properties of the liquid after degassing.

So, the biological effect of low EMF is not a phantom but has an obvious physical basis. In this model the effect depends on the doze and the accumulation of energy of field is originated by increase without heating in quantity of free water and its entropy.